\documentclass[letterpaper,twocolumn,10pt]{article}
\usepackage{usenix2019_v3}

\usepackage{tikz}
\usepackage{amsmath}

\usepackage{filecontents}
\usepackage{hyperref}
\usepackage{enumitem}
\usepackage{xspace}
\usepackage{listings}
\usepackage{multirow}
\usepackage{amssymb}

\newcommand{\tool}{TraceBin\xspace}

\newtheorem{definition}{Definition}

\newcommand{\gcc}{{\small \texttt{gcc}}\xspace}
\newcommand{\clang}{{\small \texttt{clang}}\xspace}
\newcommand{\flang}{{\small \texttt{flang}}\xspace}
\newcommand{\gfortran}{{\small \texttt{gfortran}}\xspace}
\newcommand{\objdump}{{\small \texttt{objdump}}\xspace}
\newcommand{\radare}{{\small \texttt{radare2}}\xspace}
\newcommand{\angr}{{\small \texttt{angr}}\xspace}
\newcommand{\ghidra}{{\small \texttt{ghidra}}\xspace}
\newcommand{\ida}{{\small \texttt{ida}}\xspace}

\newcommand{\gobmk}{{\small \texttt{gobmk}}\xspace}
\newcommand{\wrf}{{\small \texttt{wrf}}\xspace}
\newcommand{\cactus}{{\small \texttt{cactus}}\xspace}
\newcommand{\dealII}{{\small \texttt{dealII}}\xspace}
\newcommand{\perlbench}{{\small \texttt{perlbench}}\xspace}
\newcommand{\cam}{{\small \texttt{cam4}}\xspace}
\newcommand{\sjeng}{{\small \texttt{sjeng}}\xspace}

\newcommand{\htworef}{{\small \texttt{h264ref}}\xspace}
\newcommand{\xz}{{\small \texttt{xz}}\xspace}

\newcommand{\povray}{{\small \texttt{povray}}\xspace}
\newcommand{\imagick}{{\small \texttt{imagick}}\xspace}
\newcommand{\parest}{{\small \texttt{parest}}\xspace}
\newcommand{\omnetp}{{\small \texttt{omnetp}}\xspace}

\newcommand{\cpugcc}{{\small \texttt{cpugcc}}\xspace}

\newcommand{\Qa}{{\bf Q1}\xspace}
\newcommand{\Qb}{{\bf Q2}\xspace}
\newcommand{\Qc}{{\bf Q3}\xspace}
\newcommand{\Qd}{{\bf Q4}\xspace}
\newcommand{\Qe}{{\bf Q5}\xspace}
\newcommand{\Qf}{{\bf Q6}\xspace}

\newcommand{\resulteg}[1]{$\langle$#1$\rangle$}

\begin{document}
%-------------------------------------------------------------------------------

%don't want date printed
\date{}

\title{\Large \bf Evaluating Disassembly Errors With Only Binaries}

%for single author (just remove % characters)
\author{
{\rm Lambang Akbar Wijayadi}\\
National University of Singapore
\and
{\rm Yuancheng Jiang}\\
National University of Singapore
% copy the following lines to add more authors
\and
{\rm Roland H.C. Yap}\\
National University of Singapore
\and
{\rm Zhenkai Liang}\\
National University of Singapore
\and
{\rm Zhuohao Liu}\\
National University of Singapore
} % end author

\maketitle

%-------------------------------------------------------------------------------
\begin{abstract}
%-------------------------------------------------------------------------------
Disassemblers are crucial in the analysis and modification of binaries. In practice, disassemblers are known to have errors. Existing works showing disassembler errors rely on heuristic implementation without specific guarantees. They also assume source code and compiler toolchains for ground truth. However, the assumption of source code is contrary to typical binary scenarios where only the binary is available. In this work, we investigate an approach with minimal assumptions and a sound approach to disassembly error evaluation that does not require source code. Use of source code does not address the fundamental problem of binary disassembly and fails when only the binary exists. As far as we know, this is the first work to evaluate disassembly errors using only the binary, together with comprehensive experiments and error findings.

We propose \tool, which uses dynamic execution to find disassembly errors. \tool targets the use case where the disassembly is used in an automated fashion for security tasks on a target binary, such as static binary instrumentation, binary hardening, automated code repair, and so on, which may be affected by disassembly errors. Discovering disassembly errors in the target binary aids in reducing problems caused by such errors.
Furthermore, we are unaware of existing approaches that can evaluate errors given only a target binary, as they require source code.
Our evaluation shows \tool finds: 
(i) errors consistent with existing studies, even without source; 
(ii) disassembly errors identified as due to control flow;
(iii) new interesting errors; 
(iv) errors in non-C/C++ binaries; 
(v) errors in closed-source binaries;
and (vi) show that disassembly errors can have significant security implications using PoC attacks. Overall, our experimental results show that \tool finds many errors in existing popular disassemblers. It is also helpful in automated security tasks on (closed-source) binaries relying on disassemblers.
\end{abstract}

%-------------------------------------------------------------------------------
\section{Introduction}
\label{sec:intro}
%-------------------------------------------------------------------------------
Disassemblers play a crucial role in reverse engineering, binary analysis, binary-level security mechanisms, code repair, etc. The disassembly problem is to recover the assembly instructions corresponding to the code from a given binary. It is well known that perfect disassembly is generally impractical due to the intractability of the disassembly problem~\cite{horspool80,wartell11}. The difficulty is due to code and data being indistinguishable in Von Neumann architectures. Another difficulty is that deciding the target address of indirect control flow is also intractable; hence, whether some potential data should be disassembled as code cannot always be known \cite{wartell11}. In practice, there are further problems---x86 (and x64) is particularly difficult as instructions are variable length from 1-15 bytes long, the disassembler may need to handle overlapping instructions (including branching inside instruction bytes), etc. 

The primary use of disassemblers is when software is only available in binary form, e.g., closed-source COTS (Commercial off-the-shelf) software. Any analysis or hardening of the software would usually start with disassembly to obtain the machine-level program. In contrast, if the source is available, there would be little need for disassembly. Disassembly errors are significant since an error is tantamount to the disassembler giving a different (incorrect) assembly program.

We motivate why we want to determine disassembler errors given only a target binary. Suppose we have a closed-source binary $B$. We use a disassembler to obtain a disassembly $D$ from $B$. In one security application, we use an automated tool on $D$ to determine if a potentially sensitive function $f$ can be called from a certain module. If $f$ is not reached, we know that $B$ is safe wrt this requirement. The analyzer constructs the CFG from $D$, and there are no paths from the module to $f$, so it reports $B$ as safe. However, this happens to be a mistake, but not due to the analyzer---it occurs at the first step where the disassembler has made an error missing a call to $f$. Various tools rely on the disassembler to recover the program. For example, in the McSema~\cite{mcsema-web,mcsema-ref} binary lifter, a disassembler such as IDA Pro is used in constructing the CFG of the binary; thus, errors in the \ida disassembly affect lifting.\footnote{In McSema~\cite{mcsema-ref}, they argue that recovering the CFG is a difficult problem and one that a disassembler such as Ida Pro has spent effort in discovering.}

Another example using the same $B$ and $D$ is when we want to harden $B$ with a binary CFI (Control Flow Integrity) tool, which performs binary rewriting to add the requisite CFI checks. Using $D$, the analyzer determines that $f$ is unreachable, so it adds checks into the binary to only allow control flow transfer to reachable functions. An invalid control flow transfer is considered to be an attack and the default behavior would be to generate an exception. Now, when $B$ is run, there is no attack, but the program crashes due to a false positive in the CFI defense induced by the disassembly. In both examples, we see that the first step to understand what the binary $B$ is supposed to do, i.e. determine the underlying native code program of the binary. Given that there would not be source code for closed-source binaries, e.g., COTS (Commercial-off-the-shelf) software, determining the code is the task of the disassembler. Naturally, errors in the disassembly can compromise other security tools that rely on the disassembly, which motivates our work. Examples showing failure on security tasks are shown in \Qf Section ~\ref{sec:eval}.

In this paper, we want to evaluate errors in disassemblers given: (i) (closed-source) target binaries, i.e. no source code; and (ii) the evaluation gives a context to help explain possible reasons for an error. Given only a target binary without source code, it is not practical to find all disassembly errors in the binary. However, we show that even finding some errors is interesting and significant. 

Existing studies~\cite{andriesse16,pang21,pang22} have already shown disassembly errors to occur in disassemblers. However, these studies mainly assume the availability of a source, which means these approaches do not hold in a binary-only setting without a source. We remark that in manual reverse engineering with human-in-the-loop, some errors can be tolerated. In more automated settings, finding if a disassembly error can lead to a task failure will be helpful to improve the security process.
Our setting involves non-trivial binaries without ``human in the loop'' and mostly/fully automated workflow for scalability reasons, such as the two motivating examples. Applications such as static binary instrumentation~\cite{PEBIL,dyninst}, binary CFI~\cite{veen15,zhang13}, automated code repair~\cite{schulte13}, etc., rely on disassembly and fall in our setting. Errors in the disassembly critically affect the correctness and security guarantees of such tasks.

To find errors in disassembly tools, the key question when evaluating the disassembly is what is the correct ``ground truth''. As it is not feasible to have an always correct disassembler, i.e., an oracle that gives the correct disassembly, it is necessary to find a proxy for the oracle. Most works that evaluate the correctness of disassembly assume a heuristic strategy as the oracle, \textit{e.g.}, compiler toolchain used as an oracle. Ideally, the oracle should be {\em sound} (no disassembly error) and {\em complete} (no code missed). Existing works using heuristic oracles have shown evaluation using source code benchmarks, but do not give any guarantees of disassembly soundness/completeness. Most studies rely on a compilation toolchain \cite{pang22} using a controlled study requiring source. This {\em does not apply} in realistic disassembly use cases (see also the two motivating examples) where there is only a target binary without source code.

Our approach instead has minimal assumptions and only requires the binary, which may be stripped. It is important not to have false positives, i.e. when evaluation says a disassembled instruction is an error, it is guaranteed to be an error in the ground truth. 
The difficulty of the disassembly problem means we should not, in general, expect to identify all errors. In this paper, we explore an approach that is intrinsically sound, but has no completeness guarantees. Essentially, we pick soundness over completeness, which allows us to evaluate only with binaries. We remark that there are intrinsic limitations of the binary-only setting. We assume benign binaries (but also study trojan binaries) and not malware because we are primarily interested in disassemblers used further in security task toolchains to harden binaries. Malware is also problematic given the intrinsic theoretical difficulty of disassembly; the malware writer can exploit this to make disassembly arbitrarily difficult. The mostly benign setting is closer to binaries being produced by regular compilers, which are not adversarial.\footnote{
Existing works~\cite{andriesse16,pang21,pang22} also do not evaluate malware as they require source code.}
Similarly, self-modifying code, which need not be malware, is not in scope and also not evaluated by existing work.\footnote{It is feasible to extend our approach to deal with self-modifying code.}

In this paper, the proxy oracle that is used to evaluate a disassembler should meet the following requirements: 
\begin{description}
\item[R1: Soundness guarantees.] The oracle is sound, i.e. an error \\found by the disassembly evaluation is guaranteed to be an error.

\item[R2: Compiler and programming language agnostic.] The \\(proxy) oracle should not be tied to a compiler/programming language. This avoids restrictions on how the binary is created, e.g. should not need a particular compiler or need certain compiler options.\footnote{
Our evaluation suggests that many disassemblers are tuned to certain compilers. While this is understandable, it creates a dependency and may cause issues with changes to compilers/language standards, etc.
}

\item[R3: No reliance on source code for evaluation.] While the disassembler only works with a binary, the oracle should similarly not need any source code for the evaluation. The target binary can be closed source.
\item [R4: No debug/symbol information.] The (proxy) oracle should not need any debug/symbol information in the binaries, i.e. no restrictions on how the binary is created.
\end{description}
We highlight that these requirements are incompatible with existing approaches~\cite{andriesse16,pang21} which rely on strong source (no {\bf R3}) and toolchain 
(no {\bf R2}/{\bf R4}) assumptions (requiring source).

Our experimental evaluation shows the feasibility and usefulness of our approach, which finds many and significant disassembly errors without such assumptions. We note that Andriesse et al.~\cite{andriesse16} point out a wide range of views on difficulty of disassembly ranging from being a ``solved problem''~\cite{zhang13} to ``complex cases are rampant''~\cite{miller16}. Evaluation using a sound proxy oracle can also help shed light on these viewpoints since the disassembly problem is impractical to fully solve, and good results may be highly dependent on assumptions and other factors.

We propose to generate a ground truth (partial) disassembly which is {\em correct by construction}. Conceptually, this is achieved by simply tracing the instructions that are executed, which gives a partial disassembly of the binary. As those instructions are executed, we are certain that they should be in the disassembly. Simply using an execution trace is not be scalable. Our implementation is designed to scale to the size of the binary rather than the length of the execution trace using what we call ``unique instruction traces''. Our prototype tool, \tool, also provides an explanation feature for control flow errors and merging of unique instruction traces, which increases the scope of the proxy oracle. We evaluate several popular disassemblers (\objdump, \angr, \radare, \ghidra, and \ida), focusing on open source disassemblers but including the popular closed source \ida, to answer the following questions:
\begin{description}
\item[\Qa:] Can sound disassembly evaluation confirm existing results \cite{andriesse16,pang21}  without needing source/compiler toolchains?
\item[\Qb:] Can control flow errors be explained using the context of the source/target?
\item[\Qc:] Are there interesting new errors beyond the results of~\cite{andriesse16,pang21}? 
\item[\Qd:] As most work focused on binaries from C/C++, we investigate what happens when non-C/C++ binaries are used to evaluate disassemblers?
\item[\Qe:] Can closed-source binaries that may lack compiler-derived ground truth be used to study errors in disassemblers?
\item[\Qf:] Can trojan binaries that exploit specific disassembly errors be used to hide trojan code, i.e. instructions for an inserted vulnerability is missed in the disassembly.
\end{description}

Our experiments show \tool can be used to answer \Qa to \Qf. It has minimal requirements that allow for practical evaluation of {\bf \Qd} and {\bf \Qe}. We highlight that despite superficial similarities with existing evaluations using the SPEC CPU suite~\cite{andriesse16,pang21,pang22}, our evaluation is essentially different. This is due to a difference in the oracle employed, \cite{andriesse16,pang21,pang22} use an oracle which is closer to the ground truth but to do this requires source code and compiler toolchain support, which violates requirements {\bf R2} to {\bf R4}. Furthermore, using the compiler toolchain while giving good benchmarking results does not necessarily give soundness {\bf R1} nor completeness guarantees. Instead, we only use binaries, which is what the fundamental disassembler problem requires, and follow requirements {\bf R1} to {\bf R4}, e.g., no source code, compiler/language dependencies. Furthermore, in security tasks involving analysis/instrumentation of binaries, determining if there are any errors in the disassembly is useful since any disassembly error can make the analysis (instrumentation) to be otherwise incorrect (invalid) and {\bf Q6} shows proof of concept problems which can arise. 

In summary, we have the following contributions:
\begin{itemize}
\item
We demonstrate the feasibility of determining errors made by a disassembler given only a binary and the ability to run it. Existing works evaluating disassembler accuracy are incompatible with binary-only requirements ({\bf R3}).
\item 
The use of a sound proxy oracle guarantees that errors found are indeed errors ({\bf R1}).
\item
Experimental evaluation on many binaries firstly shows that our sound proxy oracle approach finds problems with disassemblers consistent with previous findings ({\bf Q1}) without the need for source code and new cases of disassembly errors ({\bf Q2}, {\bf Q3}).
\item
We show that disassembly errors can be significant from a security perspective. We use examples of trojanized binaries and source code with inserted (PoC) vulnerabilities to show that a disassembly error can be used to hide the inserted vulnerabilities. This shows that errors in binary disassembly can lead to attacks or bypass some defense ({\bf Q6}).
\end{itemize}

\section{Background}
\subsection{Binary Disassembly}
There are two general approaches for binary disassembly: linear sweep and recursive descent. {\em Linear sweep} essentially iterates linearly through selected regions to find machine instructions. To achieve this, heuristics can be used, \textit{e.g.,} \objdump treats invalid opcodes as an error and skips a byte~\cite{pang21}. We will mainly focus on \objdump as it is known to have good accuracy~\cite{andriesse16}. The drawback of a linear sweep is that embedded data can cause the assumption of sequential disassembly to be incorrect, as a linear sweep would mistakenly treat data as code. This causes desynchronization between what is disassembled versus the actual instructions. Fortunately, on the x86, linear disassembly is found to be self-repairing and resynchronizes~\cite{linn03}, but still, disassembly error is the consequence.

To mitigate the simple assumption behind linear sweep, i.e., code is sequentially laid out, {\em recursive descent} follows the control flow of the machine code program. Given a start address, \textit{i.e.}, the entry point of the binary, instructions are identified by following the control flow; thus, it can avoid disassembling data that lies in the text segment. However, due to its static nature, the recursive descent method follows the static control flow, which means that any indirect control flow may not be accurately identified. Various heuristics are used to discover the additional code. Pang et al.~\cite{pang21} studied the source code for various open-source disassemblers to identify the heuristics used, such as analysis using constant propagation, limited backward slicing, searching for patterns, linear sweep of code gaps, function matching using common function prologue/epilogue patterns, calling conventions for {\tt main}, symbols in the binary and unwinding information, etc. While it is clear that many sophisticated heuristics can be used, there are no guarantees given by such approaches, and they may be unsound (and incomplete). Section~\ref{sec:eval} illustrates various errors that occur in disassemblers.

\subsection{Accuracy of Disassembly}

The disassembly problem is fundamentally difficult and intractable~\cite{horspool80,wartell11}. The same difficulties arise when evaluating disassembly errors given a target binary and disassembler. As such, we should not expect to have guaranteed correct disassemblers but be willing to accept some tradeoffs in both the accuracy of the disassembler and its evaluation. Realistic assumptions are also important. We assume that only the binary is available ({\bf R3}), i.e., closed source, and no debugging information ({\bf R4}), i.e., a stripped binary.\footnote{
In practice, one could have the problem of old source code being not
compilable with current compilers, thus preventing the use of the compiler toolchain.
}

We first define notions of correctness of disassembly in our setting. Let $B$ be a binary file. We assume, for the sake of definitions, the existence of an oracle $O$ which gives the correct disassembly. The disassembly produced by disassembler $A$ on $B$ is denoted by $D_A(B)$, and the correct disassembly with the perfect ground truth oracle $O$ is $D_O(B)$. We also represent each instruction as a tuple of the machine instruction and its address.

\begin{definition}
{\em A disassembly $D_A(B)$ is {\em sound}
if $D_A(B) \subseteq D_O(B)$}. 
\end{definition}
\begin{definition}
{\em A disassembly $D_A(B)$ is {\em complete} if $D_A(B) \supseteq D_O(B)$.}  
\end{definition}

A disassembly is represented as a set.
% \footnote{
% Instructions are a tuple of a machine instruction and a location/address.
% }
Informally, {\em soundness} means no disassembled instruction is incorrect. Consider an instruction $d$ in disassembly from disassembler $A$ ($d \in D_A(B)$); the proxy oracle should be sound to avoid making a mistake in the evaluation of $A$ wrt instruction $d$, i.e., if $d$ is an error then indeed it is an error ($d \notin D_O(B)$). Conversely, if $d$ is said to be correct, it must be due to the ground truth ($d \in D_O(B)$).

Informally, {\em completeness} means that every instruction in the disassembly occurs in the ground truth but the disassembly may contain erroneous instructions not in $D_{O(B)}$, i.e., ``junk instructions''. Errors in a disassembly that is only complete can still be a problem; e.g., a static analysis may think that it is possible to do an indirect jump to the junk instructions. Ideally, the ground truth disassembly is one which is both sound and complete disassembly, i.e., $D_A(B) \equiv D_O(B)$, but this may be impractical.

The challenge of disassembly lies in the perfect ground truth disassembly oracle being practically infeasible in the worst case. As such a more limited proxy is needed. Previous studies have predominantly relied on static approaches to establish ground truth under strong assumptions. The approaches are well described by Pang et al.~\cite{pang22}, we summarize:
\begin{itemize}
\item Manual analysis of the results---This is inherently unscalable;
\item Reusing existing disassemblers---Unfortunately, the problem is that every disassembler has its own drawbacks;
\item Using intermediate compiler outputs---This assumes the compiler is the correct oracle and requires source;
\item Leveraging compiler metadata~\cite{andriesse16}---There may be limitations in the use of symbols and line metadata;
\item Tracing compiling process~\cite{pang21}---Modifying the compiler tool\-chain is not scalable.
\end{itemize}
The above approaches are either not automatable or require source code and compiler-based toolchains.  We highlight that there has been comprehensive work in~\cite{andriesse16,pang21} with extensive benchmarking, but they only help to show where disassemblers work or have errors on a fixed benchmark, but one cannot generalize those results to a given target binary. Furthermore, the source code and toolchain assumptions do not hold in a binary-only situation (requirement {\bf R3}/{\bf R4}), which is our goal. 

Existing works do not guarantee soundness or completeness; rather, their evaluation is essentially empirical. Using the compiler toolchain assumes that relying on compiler modifications and using compiler outputs is reliable because the compiler is reliable. The experimental evaluation is essentially empirical, and any conclusion only holds for the tested results, including versions, options, etc. Thus, there are inherent difficulties in providing guarantees on such heuristic oracles.

We highligh that there can still be subtle issues with using compiler toolchains. History shows that compilers do have bugs, e.g., fuzzers such as Csmith~\cite{Csmith} have successfully found compiler bugs. Debug information in the compiler toolchain has also been shown to have bugs~\cite{li2020,luna2021}. Modified compiler toolchains may not scale over time as language and compilers inevitably change, so the evaluation may be restricted to a certain version of the compiler and language. For example, a modified toolchain for C++17, e.g. based on Clang 6, may not be able to evaluate a program using C++20 even when there is both source and binary.

Notwithstanding that using compiler toolchains also have limitations, using compilers imply availability of source code.  This changes the binary problem to the source code setting, which breaks our requirements {\bf R2} to {\bf R4}. The primary reason for using disassemblers is usually because of a lack of source code. Consequently, the natural use case of disassemblers is the disassembly of (closed source) binaries, which is inherently incompatible with any assumptions requiring heuristic toolchain-based oracles such as in~\cite{andriesse16,pang21,pang22}. We show that a different approach, with a sound approach and minimal assumptions, can still provide a valuable evaluation of disassembler errors. However, our goal is not to perform comprehensive experiments on how disassemblers behave under different compiler settings; rather, we want to explore the soundness approach to see how well it works to find errors on existing disassemblers given only target binaries.

As it is not practical to obtain the ground truth, a disassembler typically makes a tradeoff between completeness and soundness, which we summarize:
\begin{itemize}
\item Sound \& Complete: This is impractical as it means that the disassembler is equivalent to the ground truth oracle $D_O(B)$. 

\item Sound \& Not Complete: This is the approach in this paper, \tool, for evaluating disassemblers.

\item Not Sound \& Not Complete: There are neither soundness nor completeness guarantees. Our evaluation shows that the recursive descent dissemblers studied (\angr, \ghidra, \radare, \ida) belong in this category, incomplete because instructions are missed (e.g., see \Qb in Section~\ref{sec:eval}) and unsound with wrong disassembly (e.g., see Appendix \ref{appendix:wrong_disasm}). 

\end{itemize} 
The evaluation of disassembly errors will also need to deal similarly with these tradeoffs, given the lack of a perfect oracle, which would rely on some form of proxy oracle.

\begin{figure}[tb]
  \centering
  \includegraphics[width=0.45\textwidth]{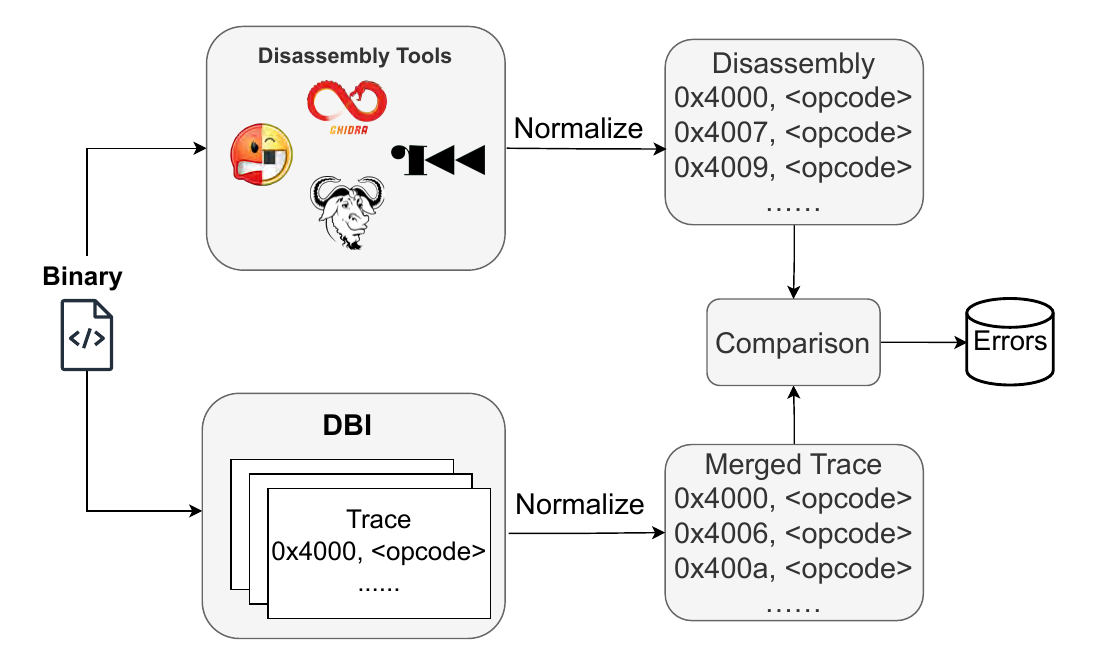}
  \caption{Overview architecture of \tool. \label{fig:arch}}
  % \Description[Overview architecture of \tool.]{Overview architecture of \tool.}
\end{figure}

\section{Design}
\label{ch:design}

The goal is to have a sound but incomplete oracle, so our evaluation of disassemblers given a target binary will not generally be complete. Ideally, there should be minimal assumptions on the binary. In order to be sound, we rely on the observation that instructions that are executed must be contained in the ground truth oracle. A proxy oracle based on execution traces is intrinsically sound. We do not claim any special novelty in this observation, but we are not aware of any (existing) approach that has soundness guarantees. Our contribution is to show that such a simple approach can answer Questions {\bf Q1} to {\bf Q6} while only working with the binaries themselves (Requirements {\bf R1} to {\bf R4}). As previously discussed, given the tradeoff between soundness and completeness, and we focus on soundness.

The basic abstraction used is as follows. A binary has instructions and data. An instruction $x$ in the binary can be considered as being loaded at an address in the process denoted by $addr(x)$. A disassembler disassembles the binary into instructions, an instruction $y$ in the disassembly would be identified to be at location $loc(y)$. Once there is a mapping between $loc(y)$ and $addr(x)$, then we can say that the disassembly is {\em correct} on instruction $y$ when $x$ and $y$ are the same instruction and are at the same address. Otherwise given $x$, $y$ is an {\em erroneous instruction} in the disassembly, and in this paper, we simply call this an {\em error}. Note that an error could occur due to either: (i) a mismatch in the machine instruction, i.e. the executed instruction at an address is different from the the instruction in the disassembly mapping to that address; or (ii) there is no mapping from $x$ to the disassembly, i.e. the disassembler has missed $x$. As errors are only found using instruction traces, we do not say anything about instructions in the disassembly that are not executed. Determining the mapping can depend on the operating systems. More details of our \tool prototype are in Section~\ref{ch:implementation}.

Generating an execution trace seems conceptually simple; however, the design needs to be scalable in order to be usable. Consider the \gcc benchmark from SPEC 2006 ({\tt 403.gcc}). Executing the ref workload on a binary compiled with {\tt gcc -O3} averages to $\approx$9.1e10 instructions
(from the binary) per input,\footnote{
A lower bound assuming an instruction takes at least one byte on average in the trace means at least 1GB of trace for {\tt 403.gcc} .
}
and runtime is $\approx$130s.
Clearly, naive collection of the trace will not scale as a pure trace can be very large. We remark that disassemblers can also take substantial runtime, e.g., \angr takes $\approx$842s disassembling {\tt 403.gcc}.\footnote{
The runtime in this case for disassembly is greater than the SPEC runtime.
}

For a more scalable design, we make the useful observation that for our purposes, we can relax $T$ so that it is no longer a sequence of instructions but rather a {\em set of instructions} where an instruction is composed of an instruction tuple consisting of the machine instruction and its address. We call this the {\em unique instruction trace}. If $T$ is a sequence of instructions, then the size of $T$ is linear in the execution time, which can be very large. To avoid the cost of dealing with an $O(T)$ sized sequence, we design the size of the unique instruction traces instead to be $O(\mbox{binary size})$. For example, the {\tt 403.gcc} binary has $\sim 690$K instructions, which is significantly smaller than the instruction trace as a sequence of $9.1e10$ instructions.

In practice, the average binary may be relatively small and both known and fixed. We regard the size of the unique instruction trace as $O(1)$ making comparing the trace with the disassembly practical. Furthermore, trace $T$ need not be from a single run but can be a set of unique instructions collected from multiple runs of the program, which we call instruction merging and can be used to increase instruction coverage. In this paper, we will also use the term trace to refer to the set of unique instructions and their addresses.

The sound proxy oracle is the collected execution trace $T$ (set of unique instructions across runs). We evaluate the errors in a disassembler by checking all instructions in $T$ for an error, \textit{i.e.}, $x\in T$ without a matching $y$ in the disassembly. As the proxy oracle is sound, a mismatch from this process is guaranteed to be an error.

There are inherent tradeoffs in identifying the correct assembly code or, conversely, the errors in the disassembly. Our approach has the advantage that it is not dependent on any compilers or programming languages, nor does it need any meta information such as symbols. From the perspective of this paper, the most important property is the sound by-construction guarantee. The limitation is that unless the combined execution traces cover all the code, this approach will be inherently incomplete. However, as guaranteeing soundness and completeness is impractical, this is a reasonable tradeoff choice. In this paper, we implement and investigate the effectiveness of the sound approach. As far as we know, this approach has not been seriously investigated.

\section{Implementation}
\label{ch:implementation}

\begin{figure}[tb]
  \centering
  \includegraphics[width=0.47\textwidth]{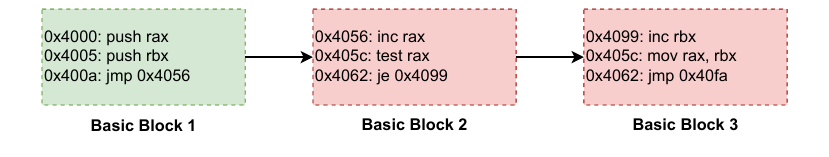}
  \caption{Explaining control flow errors 
  \label{fig:auto-expl}}
  % \Description[Explaining control flow errors.]{Explain how \tool mark control flow errors.}
\end{figure}

The implementation of \tool is designed to meet requirements {\bf R1} to {\bf R4}. It consists of three parts: (i) The binary is run to collect the trace (unique instructions executed); (ii) Disassemblers are run to collect their disassembly result; and (iii) Comparison of instructions in trace vs the disassembly. An overview of the \tool steps is given in Figure~\ref{fig:arch}. Although trace collection may seem simple, a practical tool must be efficient and scalable. We use dynamic binary instrumentation (DBI) to collect the unique instruction trace with DynamoRio~\cite{bruening12} (other DBI are feasible). Our prototype focuses on \verb+x64+ binaries on Linux.

The basic dynamic binary instrumentation checks whether a basic block has been executed before or not (we employ a hash table). Working at the basic block level is much more efficient than processing instructions one by one, {\it e.g.}, running {\small\tt blender} (SPEC 2017) takes 173s native (without DBI), 233s collecting unique instruction trace with our basic block approach, while instrumentation per single instruction level takes 40,829s. Our instrumentation is more efficient as it avoids processing the individual instructions for subsequent executions of a basic block. In our evaluation, we found that the overheads of DBI and tracing are reasonable.

Although the current implementation uses DynamoRIO, a user\-space instrumentation tool, it can be extended to support kernel space binaries. While outside the scope of this paper, one way to evaluate disassembly of kernel space binaries is to use kernel-space instrumentation tools like KProbes~\cite{kprobes06}, LTTng~\cite{desnoyers10}, or DynamoRio as a kernel module (DRK) ~\cite{drk}.

We also capture details of control flow transfer (direct and indirect) instructions during execution, recording the sets of source and destination addresses of instructions. This provides the {\em control flow explanation} feature in \tool. After finding a disassembly error, there are some possible explanations, and \tool flags potential reasons for the control flow error. First, it distinguishes between source and target control flow possibilities: (i) a source error is when the original control flow instruction is not in the disassembly, {\it i.e.}, a (partial) basic block is missed; and (ii) a target error when the control flow target of the control flow is missed, although the control flow instruction itself is correctly disassembled. Figure \ref{fig:auto-expl} illustrates these cases; consider when basic block 1 is disassembled, but basic blocks 2 and 3 are not. The jump instruction in Basic Block 1 is the source, and Basic Block 2 is the target. As basic block 2 is missed, this is case (ii) above (target error). Because the control flow within block 2 is not disassembled, basic block 3 is also not disassembled, and we consider block 3 to be case (i) above (source error). An example of case (ii), which shows the control flow error, can be seen in Appendix \ref{app:direct}. These checks are further extended to call-return control flow to determine the context of control flow errors there. We find sometimes disassemblers do not correctly disassemble the continuation of a function call after a return (see Section \ref{sec:eval} \Qb).

\tool normalizes the virtual address of the instruction to the base address captured during runtime instrumentation. More specifically, an instruction $x$ in the trace has its address $addr(x)$ adjusted by subtracting the base address of the corresponding text section of the code. This deals with differences in runtime code addresses which can occur such as through ASLR and PIC. The mapping between instruction $x$ in the trace and instruction $y$ in disassembly is then direct, i.e., $addr(x) \equiv loc(y)$.

We also support the merging of trace sets, combining multiple runs into a single trace. Merging trace sets can substantially enhance evaluation coverage, e.g., merging executions of the {\small\tt rar} binary with various options can expand the trace from 3.8K unique instructions to 22K unique instructions. 

Every disassembler has its own output format. We leverage the framework in OracleGT \cite{pang21,pang22} to simplify gathering the assembly output from each disassembler evaluated. Some disassemblers assign different base addresses, \textit{e.g.}, \ghidra adds 0x100000 to the base address while \angr adds 0x400000 for non-PIC binaries. We also normalize all these addresses in order to make a comparison with the trace. Essentially, we find disassembler errors by cross-referencing every instruction in the trace against the static disassembly to determine if there is a (disassembly) error. The limitation of our approach is that we are limited by instruction coverage from the union of merged trace sets.

We also assume that the user of \tool knows how to use and run the binary. Fuzzing techniques can also be used to increase coverage. We remark that binaries that require complex disassembly, where more than one disassembly is possible due to self-modifying code and overlapping instructions, can also be dealt with using extensions of our approach.

\section{Evaluation}
\label{sec:eval}
We evaluate \tool on \Qa to \Qf on programs from SPEC CPU 2006 and CPU 2017 by merging the {\em test, train, and ref workloads}, closed source binaries (RAR, Tigress, PNGOUT, and CUDA), and some test programs. The SPEC programs contain C, C++, and Fortran code. For C/C++ code, \gcc 9.4.0 and \clang 10.0.0 are used. We will sometimes refer to a SPEC program by its binary filename for disambiguation. For Fortran code, \gfortran 9.4.0 is used for Fortran only code, and when the code is a mix of Fortran and C, we use \gfortran \& \gcc or \flang 7.0.1 \& \clang depending on whether the evaluation is on \gcc or LLVM. Binaries are all \verb+x64+ binaries on Linux. The disassemblers evaluated are popular ones: linear sweep: \objdump, and recursive-descent: \angr, \ghidra, \radare, and \ida.\footnote{
Details of the disassembler versions are:
{\tt objdump} 2.34 (\url{https://ftp.gnu.org/gnu/binutils}),
{\tt angr} 9.2.79 (\url{https://github.com/angr/angr}),
{\tt ghidra} 10.3.1 (\url{https://ghidra-sre.org}) and 
{\tt radare} 5.8.8 (\url{https://github.com/radareorg/radare2}).
{\tt ida} 7.6 (\url{https://hex-rays.com/})
}
We focus the evaluation on open source disassemblers because they are not encumbered, but we have also added \ida given its popularity as a commercial disassembler and often said to be ``best of class''.

\begin{table*}[tb]
\centering
\caption{Overall results for SPEC 2006 \& 2017 binaries: \checkmark: no errors (Z column); \#programs with range of errors found--A: 1-80 errors, B: 81-410 errors, C: 410-978 errors, D: 1010-112351 errors; T: total number of instruction errors for all programs in A to D}
\setlength{\tabcolsep}{4pt}
{
\small
\begin{tabular}{l|c|ccccc|ccccc|ccccc|cccccc}
\hline
\multicolumn{1}{l|}{Compiler Flags} & \multicolumn{1}{c|}{objdump} & \multicolumn{5}{c|}{angr} & \multicolumn{5}{c|}{ghidra} & \multicolumn{5}{c|}{radare2} & \multicolumn{5}{c}{Ida} \\
& Z & A & B & C & D & T & A & B & C & D & T & A & B & C & D & T & A & B & C & D & T \\
\hline
gcc -O0 & \checkmark & 1 & 0 & 0 & 0 & 9 & 3 & 0 & 1 & 4 & 62855 & 4 & 5 & 4 & 13 & 242079 & 11 & 0 & 0 & 8 & 88249 \\
gcc -O0 (strip) & \checkmark & 4 & 0 & 0 & 0 & 42 & 3 & 0 & 1 & 4 & 62855 & 5 & 5 & 4 & 12 & 233383 & 21 & 3 & 0 & 8 & 89375 \\
gcc -O3 & \checkmark & 5 & 0 & 0 & 0 & 28 & 6 & 1 & 1 & 6 & 58303 & 5 & 5 & 3 & 6 & 56762 & 2 & 4 & 4 & 7 & 39635 \\
gcc -O3 (strip) & \checkmark & 7 & 0 & 0 & 0 & 52 & 6 & 2 & 0 & 6 & 57747 & 6 & 2 & 3 & 7 & 60061 & 3 & 4 & 4 & 7 & 40853 \\
gcc -O3 -flto & \checkmark & 4 & 0 & 0 & 0 & 27 & 5 & 4 & 1 & 7 & 64442 & 1 & 5 & 1 & 11 & 58421 & 5 & 3 & 0 & 10 & 48847 \\
gcc -O3 -flto (strip) & \checkmark & 10 & 0 & 0 & 0 & 133 & 5 & 4 & 1 & 7 & 64442 & 4 & 4 & 1 & 11 & 70568 & 5 & 4 & 0 & 10 & 53862 \\
clang -O0 & \checkmark & 23 & 17 & 0 & 0 & 3448 & 0 & 1 & 0 & 0 & 147 & 16 & 10 & 2 & 12 & 140951 & 23 & 17 & 0 & 0 & 3440 \\
clang -O0 (strip) & \checkmark & 23 & 17 & 0 & 0 & 3482 & 0 & 1 & 0 & 0 & 147 & 16 & 10 & 2 & 12 & 150748 & 23 & 17 & 0 & 0 & 3440 \\
clang -O3 & \checkmark & 23 & 17 & 0 & 0 & 3208 & 3 & 3 & 3 & 4 & 27470 & 19 & 9 & 5 & 7 & 66355 & 23 & 17 & 0 & 0 & 3206 \\
clang -O3 (strip) & \checkmark & 23 & 17 & 0 & 0 & 3208 & 3 & 3 & 3 & 4 & 27545 & 8 & 4 & 1 & 27 & 544840 & 19 & 16 & 0 & 5 & 25133 \\
clang -O3 -flto & \checkmark & 23 & 17 & 0 & 0 & 3208 & 5 & 4 & 3 & 6 & 47091 & 17 & 8 & 4 & 11 & 165568 & 23 & 17 & 0 & 0 & 3208 \\
clang -O3 -flto (strip) & \checkmark & 23 & 17 & 0 & 0 & 3208 & 5 & 4 & 3 & 6 & 47117 & 6 & 4 & 3 & 27 & 774288 & 17 & 17 & 1 & 5 & 34220 \\
\hline
\end{tabular}
}
\label{tab:overall_results}
\end{table*} 

As we use default options in the compilers except for the optimization level, binaries that are compiled with \gcc have PIC enabled, while those compiled with \clang do not. Experiments are run on Ubuntu 20.04. For brevity, when we say ``with \gcc'' or ``with \clang'', it is a short form for ``compiled with \gcc/\clang''.

The reason for using SPEC 2006 and 2017 is two-fold. Firstly, SPEC 2006 has been used in previous works~\cite{andriesse16,pang21}, which allows some comparisons. We highlight that although the evaluation may seem similar to~\cite{andriesse16,pang21}, it is quite different because we do not use any source code or compiler toolchain in the disassembler evaluation. On the other hand, \cite{andriesse16,pang21} require source code, which means that it does not evaluate disassemblers with only binaries and does not apply to target binaries. Additionally, we evaluate SPEC 2017.

Secondly, the SPEC programs are defined by their workload. This means that due to the nature of SPEC, only the coverage of the execution from the workload defines the SPEC code, and other parts of the code are not significant from the SPEC perspective. As such, a complete disassembly is also not needed for SPEC; instead, the relevant code is covered by the workload.

The coverage of the disassembly with \tool depends on what has been executed. With the SPEC benchmarks, the workloads are predefined as such there is a fixed coverage for the SPEC benchmark. Separate from \tool, which only collects (merged) execution traces, we have also collected the coverage for both SPEC 2006 and 2017 benchmarks combining on the workloads (test, train, and ref). The coverage is obtained by compiling the benchmarks using \gcc with turn on the \verb+coverage+ flag to instrument code for coverage analysis and using the coverage is obtained with \verb+gcov+. Note that as coverage is independent of \tool, it can be collected separately. We highlight that the purpose of the coverage statistics is simply to understand how much of the SPEC benchmarks are covered in the disassembler evaluation when SPEC has been used. Coverage of the binaries with respect to the source code is possible for SPEC benchmarks simply because source code is available. For a target binary without source code availability, it will not be feasible to have source code coverage.

On average, SPEC 2006 and 2017 achieve 47.5\% coverage. The highest coverage is observed in \texttt{mcf} 2006 at 88\%, while the lowest is in \texttt{imagick} 2017 at 9.1\%. As SPEC workloads are intended for performance evaluation, it is expected to be lower for \texttt{imagick} which is an image processing utility since it would only exercise part of the functionality of \texttt{imagick}. Still, disassembler errors are found in \texttt{imagick}.

To discuss specific results, we use the notation \resulteg{{\it Opt}+{\it Dis}, {\it Qualifier}} for flexibility and compactness where {\it Opt} is the compilation options for the binary, {\it Dis} is the disassembler evaluated and {\it Qualifier} further restricts the result. For example, \resulteg{{\tt Clang -O0} + \ghidra, {\tt gobmk}} gives the number of disassembly errors from \ghidra found with the \gobmk benchmark. The qualifier may also be a column name, \textit{e.g.}, column D in \resulteg{{\tt gcc -O0} + \radare, D}.

\noindent
{\bf Q1 (Correlation with existing studies)}
In this paper, we experiment with a sound ground truth proxy oracle approach, which differs from existing evaluations for an alternative approach. While there are comprehensive evaluations of disassembler accuracy~\cite{andriesse16,pang21,pang22}, they do not use (sound) ground truth oracles. We highlight that there is a tradeoff between soundness and completeness, so it is a choice. Here, we investigate the accuracy of disassemblers purely from a soundness perspective to complement other forms of evaluation. Given that different oracle tradeoffs are not directly comparable, we can still see how results correlate. We caution that any comparison with previous works should account for compiler and disassembler versions, and we use a larger SPEC benchmark (SPEC 2006 and 2017).

Table \ref{tab:overall_results} gives overall results for the binaries tested. Rows are binaries created under the option in column Flags. Two programs from SPEC 2017 (hybrid C \& Fortran), \wrf and \cactus, are omitted due to known compilation issues with \verb+-flto+. Pure Fortran programs are discussed separately. Altogether, there are 23 SPEC 2006 and 17 SPEC 2017 programs in Table~\ref{tab:overall_results}. Binaries are not stripped except for rows that explicitly mention it. Column Z (with \checkmark) shows no errors were found with \objdump. Columns from A to D give the number of programs where the disassembler was found to have errors within a certain range, \textit{e.g}, column A counts the number of programs giving 1-80 errors. The idea for this presentation is: (i) to make it easier to visualize the errors on a configuration (row); (ii) to show the number of programs where the disassembler has errors together with (iii) a measure of the degree of errors (A (least) to D (most)). 

To illustrate our table, it shows with \verb+gcc -O0+, eight programs (SPEC 2006: \dealII, and \povray, SPEC 2017: \perlbench, \parest, {{\small \texttt{pov\-ray}}\xspace}, \imagick, \cam, and \gcc) have disassembly errors with \ghidra, i.e., (A=3, B=0, C=1, D=4). This presentation makes it easy to see that changing from \verb+cc -O0+ to \verb+cc -O3+ gives (A=6, B=1, C=1, 6), so many more binaries now have errors (change in A-D) by increasing the optimization level, and there are also more errors per binary (not shown in table). The T (Total) column gives the total number of instruction errors found across all binaries in the 4 columns (A to D) for that row. We highlight that T is a different metric being instruction errors found while A to D are number of binaries with errors found. The Z column shows that no errors were found with \objdump (but see the later investigation into \objdump). This result is consistent with previous evaluations~\cite{andriesse16,pang21}, which report high accuracy for linear sweep disassemblers.

We highlight that our presentation of the errors on SPEC is meant to show the errors found in a disassembler on various binaries, e.g., with \verb+gcc -O0+ we see only one program is affected by errors in \angr (A=1, B=C=D=0) but with \ghidra the effect is greater (A=3, B=0, C=1, D=4) (possibly with greater consequences for these errors and across more software/binaries). The result presentation in~\cite{andriesse16,pang21} differs, focusing on error statistics as a whole, e.g., the T column. Our presentation instead is targeted towards our goal of evaluating errors with target binary, so while Table~\ref{tab:overall_results} has statistics, it focuses more on the binaries themselves and variations in binaries arising from creation options such as optimization level, etc. To summarize, the presentation of errors in~\cite{andriesse16,pang21} is totally different, focusing on aggregate errors.

We now discuss recursive descent disassemblers, and all had errors. Table~\ref{tab:overall_results} shows that \angr has the best overall performance for the recursive disassemblers on these binaries. This is because columns C~\&~D are 0 for all rows, indicating there is no large number of errors per binary for \angr. By contrast, \ghidra has a smaller A column for many rows but larger B to D columns, which suggest a larger impact with more more errors when they occur (more binaries have errors and more instruction errors). We can see that in terms of error impact across programs and compilation options, \ida has more programs having errors than \angr, and columns C~\&~D are not zero columns, i.e., there are binaries with a significant number of errors, whereas \angr has zeroes in C~\&~D. The T column (total aggregate errors) is smallest for \angr as well. We might be tempted to order performance as \angr $\succ$ \ghidra $\succ$ \ida $\succ$ \radare where $\succ$ denotes ``better''.

However, Table~\ref{tab:overall_results} is only meant to give the overall picture by aggregating the results. In the detailed results (not shown), we find that no recursive disassembler dominates all other recursive ones, i.e., no disassembler is strictly better than the others. This may be surprising since \radare seems to have the worst overall results in Table~\ref{tab:overall_results}.

Some examples showing non-dominance follow. In \gobmk SPEC 2006, \angr is worse than \ghidra: \resulteg{{\tt Clang -O0} + \ghidra, \gobmk} = 0 vs \resulteg{{\tt Clang -O0} + \angr, \gobmk} = 60. Conversely, in {\tt sjeng} SPEC 2006, \ghidra is worse than \angr: \resulteg{{\tt gcc -O3} + \angr, \sjeng} = 0 vs \resulteg{{\tt gcc -O3} + \ghidra, \sjeng} = 1331. Similarly, \perlbench SPEC 2006 shows \radare is worse than \ghidra: \resulteg{{\tt gcc -O0} + \radare, \perlbench} = 13869 vs \resulteg{{\tt gcc -O0} + \ghidra, \perlbench} = 0. However, when the same binary is compiled with \gcc \verb+-O3+, \ghidra is worse than \radare: \resulteg{{\tt gcc -O3} + \ghidra, \perlbench} = 7016 vs \resulteg{{\tt gcc -O3} + \radare, \perlbench} = 9. Turning to \ida and \omnetp 2017, \ida is worse than the other recursive descent disassemblers: \resulteg{{\tt gcc -O0} + \ida, \omnetp} = 3542 vs \resulteg{{\tt gcc -O0} + \radare, \omnetp} = 431; while \ghidra and \angr have 0 disassembly errors found on the same binary. On the other hand, \ida is better than \ghidra for \cam 2017: \resulteg{{\tt gcc -O0} + \ida, \cam} = 0 vs \resulteg{{\tt gcc -O0} + \ghidra , \cam} = 3333. We see that the detailed results give a very mixed picture showing that overall general performance cannot easily be extrapolated to specific binaries.

The example below is a disassembly from \ida of \verb+omnetpp+ 2006 compiled with \verb+gcc -O0+:
\begin{lstlisting}[basicstyle=\scriptsize]
    17b6fc: callq  17ae81
    17b701: jmpq   183f37 
    17b706: .... (not disassembled)
\end{lstlisting}
\ida does not disassemble instructions in the address range {\tt 0x17b706 - 0x17b7f3}, but those instructions are correctly disassembled by the other recursive descent disassemblers (\radare, \ghidra, and \angr) showing that in a portion of the binary, \ida gives a worse result the other disassemblers eventhough \radare is generally worse. More examples for all recursive descent disassemblers (\angr, \ida, \ghidra, and \radare) to illustrate non-dominance are given in Appendix \ref{app:nondom}.

Table~\ref{tab:overall_results} also shows errors in recursive disassemblers depending on how the binaries are created. Previous works~\cite{andriesse16,pang21} suggest that higher optimization has an effect and that higher optimization levels may result in more errors. The existing results are what one might expect. We now discuss our results, focusing on \verb+-O0+ vs \verb+-O3+ (LTO is discussed later) and present a more nuanced view. There are cases where \verb+-O0+ is worse than \verb+-O3+: (i) \resulteg{{\tt gcc -O0} + \radare, D} = 13 vs \resulteg{{\tt gcc -O3} + \radare, D} = 6 and the total errors go from around 242K down to 56K; (ii) similarly for \verb+clang -O0+ to \verb+-O3+ on \radare the total errors decrease. Thus, we see that while \verb+-O3+ can present more challenges, \verb+-O0+ also has challenges for existing recursive disassemblers.

We expect the number of errors to increase when the symbol table is unavailable, i.e., stripped binaries. Table \ref{tab:overall_results} shows the expected result in the aggregate. However, there are some counter-intuitive instances. The opposite effect, stripped has fewer errors than non-stripped, can be seen in: \resulteg{{\tt gcc -O3} + \ghidra, \imagick} = 710 while \resulteg{{\tt gcc -O3} strip + \ghidra, \imagick} = 154. There are also cases when the number of errors increases significantly when stripped, \textit{e.g.}, \resulteg{{\tt gcc-O3} + \radare, \cactus 2006} = 754 vs \resulteg{{\tt gcc -O3} strip + \radare, \cactus 2006} = 5159. We remark that this seems to be a new finding as it is not reported in previous works~\cite{andriesse16,pang21} (while this is part of our new results, it is mentioned in this section to be in the context of the Table \ref{tab:overall_results} discussion).

\noindent
{\bf \Qb (Control flow errors)}
Table~\ref{table:auto_explain_res} shows the result of disassembly errors caused by control-flow instructions in the SPEC benchmarks, focusing on the instructions in a control-flow transfer that are missed in the disassembly. The table is divided into four categories: cbr (conditional branch), indirect, direct, and return. The table shows the number of errors found using the control flow explanation feature in \tool for each disassembler. 

The number of errors is counted as the total number of instructions in the missed target basic block $B$. However, this may not account for all the control flow errors as $B$ may have further control flow transfer to another basic block $B'$. The errors due to $B'$ are not counted in this table (see Section~\ref{ch:implementation}, control flow explanation case (i)) because we do count for the transitive effect of control flow errors beyond $B$ in our implementation so that the trace collection is efficient. We account indirectly for these errors as they are included in the total errors in Table~\ref{tab:overall_results}. Examples of control flow errors found are given in Appendix~\ref{app:controlflow}.

We see the main cause of control flow errors is due to indirect jump/call---indirect control flow errors dominate others in Table~\ref{tab:overall_results}. Examples of indirect control flow errors are in Appendix~\ref{app:indirect}). Control flow errors are expected for the recursive descent disassemblers since heuristics are used to determine the target of the indirect call/jump. What is surprising is disassembly errors in \angr, \ida, and \radare for unconditional and conditional direct call/jump. We did not expect that the disassembler would have control flow errors on control flow transfers, which are direct. The example below is a result from \radare of {\tt Xalan} 2006 compiled with {\clang \tt -O0}.
\begin{lstlisting}[
    basicstyle=\scriptsize,
    label={lst:direct_err}]
    4ae3db:       callq  5d1080 
    4ae3e0:       mov    %rax,-0x2d8(%rbp)
    4ae3e7:       jmpq   4ae3ec 
    4ae3ec:       (not disassembled)
    .............................................
    4aff41:       (not disassembled)
    4aff42:       lea    -0x28(%rbp),%rdi
    4aff46:       callq  4f9980
    4aff4b:       jmpq   4aff63
\end{lstlisting}
\radare only disassembles until address {\tt 0x4ae3e7}, does not disassemble from address {\tt 0x4ae3e7}, and restarts disassembly again at address {\tt 0x4aff42}. The jump instruction at address {\tt 4ae3e7} works like no-op instruction as it just jumps to the exact next instruction. This example also shows that \clang can systematically produce redundant jumps as the target of the jump is the next instruction (at {\tt 0x4ae3ec}).
We conjecture that code which is unusual (e.g. systematic redundant jump) may have poor interaction with heuristics in disassemblers, being unexpected code.\footnote{
In this work, we do not try to understand the code within the disassembler which creates errors. Also {\tt ida} is closed source.
}

We also show that \ida can miss the target of direct call when disassembling {\tt Xalan} 2017 with {\gcc \tt -O0}, see details in Appendix \ref{app:direct}, Listing \ref{lst:ida_direct_err}. Similarly, conditional branches also have errors, e.g., SPEC 2006 {\tt omnetpp} with {\tt \gcc -O3 -flto} + stripped, the fall through of the conditional branch is missed by \angr, see Appendix~\ref{app:cbr}. Instructions following a return were also missed. On SPEC 2006 {\tt cactusADM} with {\tt \clang -O3 -flto}; despite the caller being correctly disassembled, instructions following the return are miss-identified by \ghidra, as shown in Appendix~\ref{app:return}. We remark that these examples are consistent with the disassemblers using heuristics, which may not be sound. We also highlight that the hardening of control flow can be important to CFI-style security applications, i.e., identifying and preventing undesired control flow. Evaluating a target binary on such errors will be relevant to such security applications.

\begin{table}[tb]
\centering
\caption{Control flow errors found in recursive descent disassemblers}
\label{table:auto_explain_res}
\begin{tabular}{|l|l|l|l|l|}
\hline
& \multicolumn{4}{c|}{\textbf{Disassembler}} \\
\textbf{Control Flow} & \textbf{Angr} & \textbf{Ghidra} & \textbf{Radare2} & \textbf{Ida} \\
\hline
cbr & 76 & 4 & 4786 & 0 \\
indirect & 19725 & 55685 & 192749 & 82771 \\
direct & 7 & 0 & 740 & 297 \\
return & 0 & 10 & 114 & 0 \\
\hline
\end{tabular}
\end{table}

\noindent
{\bf \Qc (New disassembler issues)}
We now discuss some interesting classes of errors found. As far as we are aware, these are not discussed in~\cite{andriesse16,pang21,pang22}.

\noindent
({\bf PLT}) We found that the Procedure Linkage Table can cause disassembly difficulties.
Binaries compiled with \clang have PLT sections that are not correctly disassembled by \angr, \ida, and \radare. The example below shows a disassembly of a typical PLT entry that has the code pattern \verb+jmp+, \verb+push+ and \verb+jmp+ if the binary is compiled with \clang.
\begin{lstlisting}[
basicstyle=\footnotesize,
label={lst:plt_clang}]
    401030:       jmpq   *0x2fe2(%rip)
    401036:       pushq  $0x0
    40103b:       jmpq   401020 <.plt>
\end{lstlisting}
The code after the first \verb+jmp+ instruction is missed.

Personal communication with authors of \cite{pang21} confirmed that OracleGT, did not collect instructions in the PLT section. This is not intended as a criticism of the toolchain approaches, as modifying a compiler can be complex, and it is difficult to confirm if any changes capture all the required information. However, this shows the difficulty of getting guarantees with compiler toolchain approaches, and their proxy oracle may be incomplete.

\noindent
({\bf Instructions after function epilogue})
We found cases where \ghidra and \ida failed to detect certain instructions after an instruction pattern similar to the function epilogue. We found this with SPEC 2017 {\tt xz} compiled with \gcc{} {\tt -O3 -flto}. The \verb+ret+ instruction, together with several instructions before it, may be considered to be part of a function epilogue. It appears that disassemblers may use the heuristic of stopping disassembly once a function epilogue is reached, but in general, it is still possible that there can be code after the return. 

The example below is a disassembly of {\tt xz} by \ghidra and \ida of \xz 2017 compiled with {\tt \gcc\xspace -03 -flto}.
\begin{lstlisting}[
basicstyle=\footnotesize, label={lst:ghidra_missed}]
    106f1: pop    %r14
    106f3: pop    %r15
    106f5: retq
    106f6: ... (not disassembled)
    10ae0: endbr64
\end{lstlisting}

Firstly, the address \verb+0x106f6+ is not disassembled. As no instruction at that address was executed, \tool does not say anything about whether there is a correct or erroneous instruction at that address. In the trace, instructions at addresses \verb+0x10700+- \verb+0x10acf+ are executed but is not shown in the disassembly, hence are disassembly errors in \ghidra and \ida. Basically \ghidra and \ida miss instructions from addresses \verb+0x10700+- \verb+0x10acf+ until it continues with the \verb+endbr+ instruction. 

We observe that the instructions at \verb+0x106f1+- \verb+0x106f5+ look like a function epilogue. This could be interacting with a heuristic in the disassemblers which may assume there are no instructions after what looks like a function return. We did not observe this particular problem in \radare, \angr, and \objdump. 

\noindent
({\bf O3 + LTO})
Link Time Optimizations (LTO) is a form of whole program or global optimization in \gcc and \clang compilers, which can significantly change the structure of code. Table \ref{tab:overall_results} shows that \verb+-O3+ with the addition of LTO gives significant changes in the disassembly errors. In some cases, the number of affected binaries decreases, but the total number of instruction errors is worse (see the T column). We give a \gcc example result. Evaluating \htworef from SPEC 2006, we get \resulteg{{\tt gcc -O3} + \radare, \htworef} = 276 and \resulteg{{\tt gcc -O3 -flto} + \radare, \htworef}= 6176, so turning on LTO causes many more errors.

\noindent
({\bf CET and control flow})
Intel Control-flow Enforcement Technology (CET) is Intel's hardware-based Control Flow Integrity (CFI) defence~\cite{CET}. A simple explanation of the CET hardware CFI is that the control flow target of a call or jump instruction must be a special \verb+endbr+ instruction. A binary that is compiled with CET enabled will have more instructions in it due to the additional \verb+endbr+ instructions. Given that \verb+endbr+ always denotes the call/jump target, we may expect that it might be used as part of a heuristic to find more instructions. In particular, a recursive descent disassembler can easily expand code coverage with a heuristic that searches for the \verb+endbr+ instruction, perhaps with other conditions.

To evaluate the effect of CET, we create a small test program that uses a jump table, see Appendix~\ref{app:jump}. With \gcc we can turn on CET with {\tt -fcf-protection=full} or disable the use of CET with {\tt -fcf-protection=none} compiler options. Table~\ref{table:cet_comparison} gives the number of errors on the test program with CET on or off. \radare does not disassemble the jump table portion of the binary when CET is not used in the compiled instructions but does disassemble it correctly if the CET instructions are compiled in. This suggests \radare uses a similar heuristic to what we proposed with CET. In \ghidra and \ida, the indirect jump targets are not found even when CET is enabled. There are more missed instructions in \ghidra and \ida with CET because there are more instructions in the CET binary. The fact that \angr has no errors could potentially be due to \angr also incorporating a linear sweep heuristic together with recursive descent. Finally, \objdump had no errors as the instructions happen to be contiguous.

Apart from investigating the CET heuristic, the example code in Appendix \ref{app:jump} can also be used for testing the indirect control flow errors in recursive descent disassemblers. Without the CET instructions, all recursive descent disassemblers failed to disassemble the target of the indirect jump as all of them have the same number of errors (14 errors); these 14 instructions correspond to the target of indirect jumps in lines 15-17 of Listing \ref{lst:handwrittencode}, Appendix \ref{app:jump}. Such an error can also be confirmed by \tool by using the control flow explanation feature.

\begin{table}[tb]
\centering
\caption{Comparison of Disassembly Errors With and Without CET}
\label{table:cet_comparison}
\begin{tabular}{lcc}
\hline
 & With CET & Without CET \\
\hline
Ghidra  & 19 & 14 \\
Radare  & 0 & 14 \\
Ida     & 17  & 14  \\
Angr    & 0 & 0 \\
Objdump & 0 & 0 \\
\hline
\end{tabular}
\end{table}

\noindent
{\bf \Qd (Non C/C++ binaries)} 
Apart from programs that consist of C or C++, we also evaluate Fortran programs, as \tool is agnostic to the compiler and programming language. Thus, this experiment is to show that \tool can be used on Fortran binaries to evaluate disassembly errors, illustrating Requirement {\bf R2} (compiler/language agnostic). We use the \gfortran 9.4.0 and \flang 7.0 compilers with {\tt -O0} and {\tt -O3} to create the binaries. We evaluated nine binaries that contained Fortran code in SPEC 2006 and SPEC 2017. Table \ref{table:fortran_result} shows the number of errors found for various Fortran binaries. There is no column for \objdump because we found no errors with it. We see that Fortran also introduces some errors in disassemblers. Interestingly, errors are mostly found in \radare for some of the Fortran programs. The errors in \angr and \ida are only found in binaries compiled by \flang. This is due to the issue discussed earlier, which was missed instructions in the PLT section. There is also no dominance in the recursive disassemblers but perhaps \ghidra may be considered to have an edge with fewer errors. 

\begin{table}[h!]
\small
\centering
\caption{Disassembly Results for Fortran.}
\label{table:fortran_result}
\begin{tabular}{|p{1.4cm}|l|p{0.6cm}|p{0.9cm}|p{1cm}|l|}
\hline
\multicolumn{2}{|c|}{\textbf{Benchmark}} & \textbf{Angr} & \textbf{Ghidra} & \textbf{Radare2} & \textbf{Ida} \\ 
\hline
\multicolumn{6}{|c|}{\textbf{{\tt gcc -O0} Optimization}} \\
\hline
\multirow{5}{*}{SPEC 2006} & gamess & 0 & 0 & 1475 & 0 \\
& bwaves & 0 & 0 & 0 & 0\\
& leslie3d & 0 & 0 & 9624 & 0 \\
& zeusmp & 0 & 0 & 0 & 0 \\
& GemsFDTD & 0 & 0 & 32537 & 0\\
& tonto & 0 & 0 & 769 & 0\\
\hline
\multirow{4}{*}{SPEC 2017} & fotonik3d & 0 & 0 & 10116 & 0 \\
& bwaves & 0 & 0 & 0  & 0 \\
& roms & 0 & 0 & 22998 & 0 \\
& exchange & 0 & 0 & 0 & 0 \\
\hline
\multicolumn{6}{|c|}{\textbf{{\tt gcc -O3} Optimization}} \\
\hline
\multirow{5}{*}{SPEC 2006} & gamess & 0 & 16083 & 1109 & 0 \\
& bwaves & 0 & 0 & 0 & 0\\
& leslie3d & 0 & 0 & 0 & 0\\
& zeusmp & 0 & 0 & 0 & 0\\
& GemsFDTD & 0 & 0 & 20567 & 0 \\
& tonto & 0 & 0 & 1316 & 0\\
\hline
\multirow{4}{*}{SPEC 2017} & fotonik3d & 0 & 0 & 1316 & 0 \\
& bwaves & 0 & 0 & 0 & 0 \\
& roms & 0 & 430 & 2402 & 0 \\
& exchange & 0 & 0 & 403 & 0\\
\hline
\multicolumn{6}{|c|}{\textbf{{\tt clang -O0} Optimization}} \\
\hline
\multirow{5}{*}{SPEC 2006} & gamess & 100 & 24 & 4531 & 100 \\
& bwaves & 34 & 0 & 36 & 34\\
& leslie3d & 56 & 0 & 58 & 56\\
& zeusmp & 38 & 0 & 40 & 43\\
& GemsFDTD & 134 & 0 & 13420 & 134\\
& tonto & 174 & 0 & 176 & 174\\
\hline
\multirow{4}{*}{SPEC 2017} & fotonik3d & 64 & 0 & 10182 & 64 \\
& bwaves & 40 & 0 & 42 & 40\\
& roms & 78 & 0 & 23078 & 78\\
& exchange & 68 & 0 & 70 & 68\\
\hline
\multicolumn{6}{|c|}{\textbf{{\tt clang -O3} Optimization}} \\
\hline
\multirow{5}{*}{SPEC 2006} & gamess & 102 & 204 & 352 & 102 \\
& bwaves  & 34 & 0 & 36 & 36\\
& leslie3d  & 54 & 0 & 56 & 54\\
& zeusmp  & 40 & 0 & 42 & 40\\
& GemsFDTD  & 136 & 0 & 15464 & 136\\
& tonto & 172 & 0 & 174 & 172\\
\hline
\multirow{4}{*}{SPEC 2017} & fotonik3d & 64 & 0 & 1382 & 64 \\
& bwaves  & 36 & 0 & 38 & 36\\
& roms  & 82 & 430 & 2486 & 82\\
& exchange & 64 & 0 & 475 & 64\\
\hline
\end{tabular}
\end{table}

\noindent
{\bf \Qe (Closed source binaries)}
Previously, we evaluated binaries built by compiling from source code with various compilers and options. This is precisely so that we can then compare with existing works \cite{andriesse16,pang21}, which use the SPEC benchmarks, but they use source, and we do not. The main use case for disassemblers is when only the binary is available, i.e., tantamount to closed source binaries. Since \tool has no compiler dependence, we can use it to evaluate the accuracy of disassemblers where only the binary is available. Still, the compiler-based approaches for the ground truth oracle~\cite{andriesse16,pang21} are simply not applicable.

We evaluate binaries from the following software: RAR 7.0~\cite{rar} (a popular archiver) with {\tt rar} and {\tt unrar} binaries, Tigress 3.1 ~\cite{tigress} with the {\tt cilly.native} binary, CUDA Binary Utilities ~\cite{cuda} 12.4 with the {\tt nvdisasm} binary, and PNGOUT 20200115 ~\cite{pngout} with the {\tt pngout} binary. We chose Tigress because it is an academic obfuscator that works by C to C transformation. It is deliberately distributed only in binary form. We thought it would be interesting to test an obfuscator.

The workloads used with \tool are the following. We executed {\tt rar} and {\tt unrar} using a range of operations, including compressing a directory, listing the contents of the archive, and extracting the archive. {\tt cilly.native} uses the supplied test in its distribution. {\tt nvdisasm} is used to disassemble the cubin (CUDA binary) file that is compiled from the cuda code sample. {\tt pngout} is used to optimize a png sample file. Table \ref{table:closed_source} shows the results. Interestingly, no errors were found with \objdump, which suggests no inline assembly code in those binaries. All the recursive disassemblers had errors. As with previous results, we see that \ida does not have superior results.

\begin{table}[tb]
\centering
\caption{\tool Result For Closed Source Binaries}
\label{table:closed_source}
\begin{tabular}{|l|l|l|l|l|}
\hline
\textbf{Tool} & \textbf{Angr} & \textbf{Ghidra} & \textbf{Radare2} & \textbf{Ida} \\
\hline
rar & 172 & 0 & 906 & 105 \\
unrar & 146 & 0 & 265 & 146 \\
cilly.native & 96 & 2876 & 3546 & 12 \\
nvdisasm & 104 & 0 & 10434 & 113 \\
pngout & 0 & 0 & 41 & 0 \\
\hline
\end{tabular}
\end{table}

\noindent
{\bf \Qf (Security implications of disassembler errors)}
\label{subsec:q6}
We now investigate if a disassembler error can lead to a security failure, expanding on the two motivating examples from Section~\ref{sec:intro}. For this, we will use trojanized binaries with implanted vulnerabilities. A security failure will be if the disassembler fails to disassemble the vulnerabilities, which may mean that the trojaned binary may be viewed as being equivalent to the original benign binary.

Consider a binary $B$, the goal of an attacker may be to present a trojan form of $B$, namely $B'$ containing an implanted vulnerability. $B'$ is crafted so that the vulnerability is not disassembled by disassembler $A$ by leveraging on specific errors occurring in the disassembly of $B$. We present two different attacks which give a proof-of-concept (PoC) using this idea. Firstly, we find various disassembly errors in $B$ when disassembler $A$ is used. Then, we either patch $B$ to $B'$ with binary patching or, alternatively, modify the source and recompile to obtain $B'$. The trojan binary $B'$ now has deliberately introduced vulnerabilities. To determine if the PoC is successful or not, we check that in the disassembly of $B'$ with disassembler $A$ to see whether or not the vulnerability is present. Our two PoCs follow: (details are in Appendix~\ref{appendix:attack})
\begin{description}
\item[I] Direct code patching of {\tt cpugcc} 2017 to cause an exception using the \verb+ud2+ instruction. By using desynchronization, we can trick the disassembler using linear sweep (\angr and \objdump). It also happens that this code region is also missed by \radare and \ghidra because it is called from an indirect jump. Therefore, the patched \verb+ud2+ instruction is not found by four disassemblers (\objdump, \angr, \radare, and \ghidra), i.e., the crashing behavior is not possible in the binary imagined by the disassembler, but it occurs in the ground truth. Furthermore, \tool confirms that the \verb+ud2+ instruction is executed, as shown in Listing \ref{lst:cpugcc_tracebin} Appendix \ref{appendix:attack}. A similar strategy can be applied to confuse \ida on the \xz 2017 binary.\footnote{
\xz was itself the target of a supply chain attack recently (\url{https://www.synopsys.com/blogs/software-security/xz-utils-backdoor-supply-chain-attack.html}).} We modified \xz from SPEC 2017 binary compiled with {\tt gcc -O0} to demonstrate that disassembly error in \ida can also be trojanized in a similar way to \ref{app:poc-a} and \ref{app:poc-b}. The disassembly result for \xz 2017 {\tt gcc -O0} from \ida follows:

\begin{lstlisting}[
    basicstyle=\scriptsize,
    label={lst:trojan_ida_xz}]
    19861: 48 01 d0 add    %rdx,%rax
    19864: 3e ff e0 jmpq *%rax
    19867: .....    (not disassembled)               
    19878: .....
    1987f: .....
\end{lstlisting}
We see that it does not disassemble instructions after address {\tt 0x19864} to address {\tt 0x1987f}. 

We change the instruction at address {\tt 0x19687f} to {\tt INT3}, which will introduce a breakpoint when executed. The result from \tool can be seen in the code below, which confirms that the {\tt INT3} instruction is executed; however, it is not disassembled by \ida.

\begin{lstlisting}[
    basicstyle=\scriptsize,
    label={lst:tool_xz_res}]
    0x19861: 48 01 d0               add    %rdx,%rax
    0x19864: 3e ff e0               jmpq *%rax
    0x19878: 80 bd 2f ff ff ff 00   cmpb   $0x0,-0xd1(%rbp)
    0x1987f: cc                     int3
\end{lstlisting}

\item[II] Patching at the C++ source code level of {\tt Xalan} 2006. We add a new global variable. The patch contains one vulnerability with data corruption of the global variable. We also add a new function (\verb+bad()+) and a vulnerability which calls \verb+bad()+. After the patched source is recompiled, the new binary $B'$ has the two introduced vulnerabilities. This is also confirmed by \tool. However, \radare misses both vulnerabilities in the disassembly: (i) the data corruption exploit, i.e., modification of global variable, and (ii) the call to \verb+bad()+. Unfortunately, both vulnerabilities are indeed executed but not disassembled. 
\end{description}

These examples show that a stealthy trojan binary $B'$, which is very similar to the benign binary $B$ with implanted vulnerabilities not found in the disassembly, can be systematically constructed using \tool. Note that as this is a binary attack, evaluation techniques which rely on source will not apply. In cases where the disassembly is used for a security task such as control flow analysis (example I), taint tracking (example II, variable modification), and control flow integrity (example II, call to \verb+bad+), a binary could be modified to escape the application of a hardening defense built using the disassembly. This shows that in binary security applications, which often rely on disassembly as the initial step, an evaluation of errors in the disassembly may be needed. This can be useful even if not all errors in the disassembly can be found. 

We remark that a disassembly error is not necessarily a bug in a disassembler simply because we do not expect the disassembler to be both sound and complete. So, while a disassembly error could be due to a specific bug, it could also be a {\em feature}, meaning it is a consequence of the disassembler being neither sound nor complete. It should also be obvious that these PoC examples may also bypass the manual reverse engineer as they would have to recognize that the vulnerabilities are not being disassembled. For example, \cpugcc compiled with {\tt gcc -O3} has 1.9M instructions. 

\noindent
{\bf Investigating \objdump}
Although the overall result indicates that objdump is the most accurate tool when evaluated with \tool, it is well known that the linear sweep algorithm can be tricked by inserting appropriate ``data in code'' into the text segment, resulting in desynchronization of the instruction stream leading to errors. We investigate desynchronization using inline assembly. The example code is given in Appendix~\ref{app:dataincode}, Listing \ref{lst:data_code}. We found that \objdump is indeed easily desynchronized using that code. The desynchronization results in completely different effects and instructions, including incorrect control flow, i.e., a return instruction when there is none in the ground truth, see Appendix Listing~\ref{lst:real_inst} (ground truth) and Listing~\ref{lst:obj_result} (incorrect disassembly by \objdump and \angr). We remark that this test assembly code fragment is designed to have only local dependencies, but it can easily be inserted almost anywhere in C code as it is semantically equivalent to a nop.

In addition, we found that \angr, although being based on recursive descent, was also desynchronized by this example. This may be due to \angr having a linear sweep heuristic to find code gaps~\cite{pang21}. This test found no errors with \ghidra, \ida, and \radare. We highlight this was only meant as a trivial test of desynchronization, and it also shows that \ida, \ghidra, and \radare worked for this example.

Besides this example, we show that padding instructions within a binary can be utilized to cause desynchronization, which does not disturb the runtime behavior of the binary. In the example in listing \ref{lst:cpugcc_attack_lin} in Appendix \ref{app:poc-a}, the instruction at address {\tt 0x3008d} is a padding instruction. After the binary is patched, the linear sweep disassemblers will desynchronize and fail to disassemble the instruction at {\tt 0x30090}, which will later be executed by the binary.

We remark that although \objdump appears to work well in most of the evaluation, except this one, it may not be a replacement for the proxy oracle. It is well known that the drawback of linear sweep is the confusion of data in code, which is also what we have shown. So the reliability of \objdump will depend on whether a compiler tries not to mix data inside the code. Whether or not this is always possible may depend on the instruction set of the machine. Then there is a question of whether the optimization objective is to optimize to reduce code space or runtime. Also, the compiler need not try to avoid data mixed with code, and  there may be  little control over how the binary is produced. In a controlled setting, the experimental setup is different as one can try to avoid  occurrences of data in code. Finally disassemblers based on recursive descent heuristics exist precisely because the limitations of linear sweep are well known.

\section{Conclusion}
Most works to evaluate disassemblers use some practical notion of the ground truth oracle of the binary in order to compare with the disassembly result. There are intrinsic tradeoffs, given that disassembly is not practical to solve in the worst case. As such, there is a tradeoff essentially between soundness and completeness. Most works focus on enlarging the set of disassembly errors found but may admit errors in the process given the chosen notion of ground truth. A more serious problem is that most works, in particular the comprehensive studies in~\cite{andriesse16,pang21,pang22}, employ the compiler toolchain and require the use of source code to evaluate disassembly errors. Clearly, while source code can be used in a controlled experiment setting, it cannot help in the binary-only setting, which is the main use case for disassemblers.

In this paper, we investigate a simple alternative and explore the tradeoffs given the difficulty of the disassembly problem with only binaries. We use an incomplete but sound proxy oracle to evaluate the disassembler. While this seems simple, it has the advantage of a correct-by-construction approach. With the theoretical and practical difficulties of imperfect oracles, it is unclear what other approach can guarantee soundness. Existing disassemblers already used unsound and incomplete approaches, but existing evaluations of disassemblers use techniques that require source, rendering them inapplicable in general.

We build a prototype tool, \tool, which uses dynamic binary instrumentation to gather a unique instruction trace for comparison with disassembly results from a disassembler. \tool is also more sophisticated than a basic instruction tracer, incorporating optimizations. explanation, and unique instruction trace merging features. Although our prototype is implemented with the DynamoRio DBI framework, any DBI can, in principle, be used. The limitation due to the binary setting is that our approach only guarantees soundness and thus may not guarantee complete coverage of the all the assembly code in a binary. 

Our evaluation shows that \tool can usefully answer questions {\bf Q1} to {\bf Q6} while adhering to our requirements {\bf R1} to {\bf R4}. In particular, goal {\bf R3}, no source code assumptions. We are not aware of any similar work that does not use source code and compiler toolchains together with substantial experimental results. We have taken the design choice that out of the two possible guarantees, soundness and completeness, that soundness is the choice to pick. (It is unclear how to implement a complete approach with only binaries). We also believe that tools which have guarantees are just as important as empirical tools that do not have guarantees. \tool gives soundness guarantees that disassembly errors are indeed errors, only requiring binaries with minimal assumptions. The evaluation shows its effectiveness and also gives interesting results on errors in popular disassemblers.

We evaluate \tool on binaries from SPEC 2006 and 2017 built under various options, closed source binaries, and synthetic benchmark. We show experimentally that \tool can: (i) give results that are consistent with existing works analyzing disassembly errors but without the use of source code;
(ii) find causes of errors from control flow instructions;
(iii) find interesting errors in the studied and popular disassemblers (\objdump, \angr, \ghidra, \radare, \ida);
(iv) find errors in Fortran binaries;
(v) find errors in closed-source binaries;
(vi) shows the security implications of disassembly errors. We illustrate by leveraging specific disassembly errors to generate a PoC attack to hide vulnerabilities from the disassembler with trojan binaries.

These results show that a soundness-based approach is a useful tool when disassemblers are used as part of an automated security solution starting from a binary. It can also be relevant in reverse engineering, though it is not specially designed for that setting. While there has been a large body of work both on investigating disassemblers, there seems to be a lack of work for the binary-only usage use case and non-manual uses of binary disassembly, which may require evaluation of how good the security solution built on top of disassembly.

Finally, while we expect that any heuristic chosen by the disassembler may make errors given the intrinsic difficulties of disassembly, the binaries selected are not malware and not specially constructed to be worst case. Indeed, they may be considered ``typical''. It was still a little surprising that we expected a better result from \ida, given that it is a commercial closed-source disassembler and is reputed to be among the best. Still, we did not find it to outperform open-source disassemblers in our evaluation, notwithstanding that generalizations from experimental testing are difficult to make.

\section*{Acknowledgments}
% We acknowledge the support of grant MOE-000460-01.
We thank the shepherd and anonymous reviewers for their constructive feedback. 
This research is supported by grant MOE-000460-01 and by the National Research Foundation, Singapore, and Cyber Security Agency of Singapore under its National Cybersecurity R\&D Programme (Fuzz Testing <NRF-NCR25-Fuzz-0001>). Any opinions, findings and conclusions or recommendations expressed in this material are those of the author(s) and do not reflect the views of National Research Foundation, Singapore and Cyber Security Agency of Singapore.
% and members of the CURIOSITY lab at NUS for their helpful feedback and discussions.
% \end{acks}

% \bibliographystyle{ACM-Reference-Format}


\appendix

\section{Jump Tables}
\label{app:jump}

We evaluate the use of Intel CET instructions as a heuristic for recursive descent disassemblers with the simple code in Listing \ref{lst:handwrittencode}.

\begin{lstlisting}[caption={Jump table for CET benchmark}, label={lst:handwrittencode},
basicstyle=\footnotesize,
showstringspaces=false]
1. int main()
2. {
3.    void * table[] = {&&t1, &&t2, &&t3};
4.    int idx=0, order[] = {1, 0, 2};
5.    
6.    loop:;
7.        void *addr = table[order[idx]];
8.        goto *addr;
9.    loop_end:
10.        
11.        ++idx;
12.        if (idx > 2) exit(0);
13.        goto loop;
14.    
15.    t1: printf("loc 1\n"); goto loop_end;
16.    t2: printf("loc 2\n"); goto loop_end;
17.    t3: printf("loc 3\n"); goto loop_end;
18. }
\end{lstlisting}

\section{Data in Code }
\label{app:dataincode}
The code in Listing \ref{lst:data_code} is injectable in multiple locations into source code to test if a simple ``data in code'' sequence can confuse the disassembler. Listing \ref{lst:real_inst} explains the code, giving the instructions that are executed at runtime. Listing \ref{lst:obj_result} shows that \objdump and \angr does not disassemble the \verb+incl+, \verb+nop+, and \verb+decl+ instructions.

\begin{lstlisting}[
basicstyle=\footnotesize,
caption={Example of data in code causing desynchronization}, label={lst:data_code}]
1.    jmp Label+4
2.    Label:
3.    .byte 0xe8, 0xc5, 0xfe, 0xff, 0xff
4.    .byte 0x45 .byte 0x08
5.    nop
6.   .byte 0xff, 0x4d, 0x08

\end{lstlisting}

\begin{lstlisting}[
basicstyle=\footnotesize,
caption={Executed instructions}, label={lst:real_inst}]
1.    jmp Label+4
2.    Label:
3.    .byte 0xe8, 0xc5, 0xfe, 0xff -- not executed
4.    incl   0x8(%rbp)
5.    nop
6.    decl   0x8(%rbp)

\end{lstlisting}

\begin{lstlisting}[
basicstyle=\footnotesize,
caption={objdump and angr disassembly result}, label={lst:obj_result}]
1193 <Label>:
1193: e8c5feffff        callq 105d 
1198: 450890ff4d0890    or %r10b,-0x6ff7b201(%r8)
119f: 5d                pop %rbp
11a0: c3                retq
\end{lstlisting}

\section{Examples showing Non-dominance of Recursive Descent Disassemblers}
\label{app:nondom}
We give examples to show that no recursive disassembler evaluated {\em strictly dominates the others} in terms of accuracy of disassembly. Examples of various disassembly errors are also given.

Listing \ref{lst:angr_missed} show instructions from {\tt h264} 2006 compiled with \verb+gcc+ \verb+-O0+. We found \angr misses instructions from address {\tt 0xac1b5} - {\tt 0xac1bf} but these not missed by \ida, \radare, and \ghidra.

\begin{lstlisting}[
basicstyle=\footnotesize,
caption={\angr missed instructions}, label={lst:angr_missed}]
ac1b5: jmp    ac1ca 
ac1b7: mov    -0x14(%rbp),%eax
ac1ba: mov    %eax,%edx
ac1bc: shr    $0x1f,%edx
ac1bf: add    %edx,%eax

\end{lstlisting}

Listing \ref{lst:radare_missed} gives instructions from \verb+gobmk+ 2006 compiled with \verb+gcc -O0+. \radare misses instructions from address {\tt 0x100275 - 0x1002db}, but these are not missed by \ida, \ghidra, and \angr.
\begin{lstlisting}[
basicstyle=\footnotesize,
caption={Radare2 missed instructions}, label={lst:radare_missed}]
10026f: add    %rdx,%rax
100272: notrack jmpq *%rax
100275: mov    -0x66c(%rbp),%eax
.....
1002ce: lea    0x516d33(%rip),%rdx
1002d5: mov    (%rax,%rdx,1),%eax
1002d8: cmp    $0x1,%eax
1002db: jle    100424
\end{lstlisting}

\section{Turning a Disassembly Error into a Trojan Binary with PoC vulnerabilities} \label{appendix:attack}
We present two proof-of-concept (PoC) examples turning a disassembly error into trojaned version of the binary where the disassembler misses an inserted vulnerability.

\subsection{Trojanized {\tt cpugcc}}
\label{app:poc-a}
We found disassembly errors in SPEC 2017 {\tt cpugcc} compiled with \verb+gcc -O3+ when disassembled with \radare and \ghidra. From an error found by \tool, {\tt cpugcc} executes the instruction at address {\tt 0x380090}. However, this address is not disassembled by \radare and \ghidra possibly because it is a target of an indirect jump. The instructions disassembled by \radare and \ghidra follow:

\begin{lstlisting}[
basicstyle=\footnotesize,
caption={radare and ghidra disassembly for cpugcc}, label={lst:cpugcc_attack_rec}]
38008a: 41 5c                   pop    %r12
38008c: c3                      retq
38008d: ....                    (not disassembled)
\end{lstlisting}

We also intentionally create desynchronization to confuse linear sweep disassembler (\angr and \objdump). This done by patching several bytes (\verb+e8 1f 00 0f 0b+) starting at address 0x3008d. This modification changes the bytes at address 0x380090 to an \verb+ud2+ instruction (0f 0b). The desynchronization issue in \objdump and \angr can be seen in Listing~\ref{lst:cpugcc_attack_lin}, which misses address 0x30090, i.e., \verb+ud2+ instruction does not appear in the disassembly.
\begin{lstlisting}[
basicstyle=\footnotesize,
caption={objdump and angr instructions for patched cpugcc}, label={lst:cpugcc_attack_lin}]
38008a: 41 5c                   pop    %r12
38008c: c3                      retq
38008d: e8 1f 00 0f 0b          callq  b4700b1 
380092: 2a 3e                   sub    (%rsi),%bh
380094: 79 00                   jns    380096 
\end{lstlisting}

The unique instruction trace capture with \tool shown below in Listing~\ref{lst:cpugcc_tracebin} shows the execution of the \verb+ud2+ followed by the execution of an exception handler.

\begin{lstlisting}[
basicstyle=\footnotesize,
caption={Trace result from \tool}, label={lst:cpugcc_tracebin}
]
....
0x380040,add rax,rdx
0x380043,jmp rax
0x380090,ud2 // exception instruction not found in 
             // objdump, angr, radare2, ghidra
0x85a7a0,endbr64 // signal handler
0x85a7a4,push rbp
....
\end{lstlisting}

\subsection{Trojanized {\tt xalancbmk}}
\label{app:poc-b}
\radare has a disassembly error around address 0x4ae3ec on {\tt Xalan} 2006 when compiled using \verb+clang -O0+. We identified that the missed instruction is in function {\tt expandRegis\-tryToFullSchemaSet} in file \verb+DataTypeValidatorFactory.cpp+. Injecting code into this function is undetected by \radare as shown below.

We create a variant of {\tt Xalan} modifying the source code. The ``trojaned'' version creates a vulnerable version of the code seeded with two PoC vulnerabilities where the vulnerabilities are not seen in the disassembly: (i) we add a new global variable \verb+global_var1234+, initialized to 0. The PoC is to corrupt to \verb+global_var1234+ 0x434343 at line 641; and (ii) we create a new function \verb+bad()+ and the PoC vulnerability is to call \verb+bad()+. The modified source code is in Listing~\ref{lst:xalan_inj}.

\begin{lstlisting}[
    basicstyle=\footnotesize,
    caption={Xalan injected code at 641 \& 642}, label={lst:xalan_inj}]
636: createDatatypeValidator(
       SchemaSymbols::fgDT_LANGUAGE,
637: getDatatypeValidator(SchemaSymbols::fgDT_TOKEN),
638:   facets, 0, false, 0, false); 
639: // disassembled to line 640
640: facets = new RefHashTableOf<KVStringPair>(3); 
641: global_var1234 = 0x434343; 
     // inserted vulnerability
642: bad(); // inserted vulnerability
643: facets->put((void*) 
       SchemaSymbols::fgELT_FRACTIONDIGITS,
       new KVStringPair(
         SchemaSymbols::fgELT_FRACTIONDIGITS, 
         fgValueZero));
\end{lstlisting}

\radare only manages to partially disassemble line 640.
Listing~\ref{lst:xalan_inj_disasm} shows the disassembly of the binary compiled from the modified \verb+Xalan+. \radare does not disassemble the instructions labelled with an asterisk. 
The asterisked lines do not appear in the disassembly, but are shown below for the purposes of explanation. The initial error in the disassembly leads to more errors introduced in the PoC. Specifically, the PoC modifies variable \verb+global_var1234+ and makes a call to \verb+bad()+ but both cannot be seen in the \radare disassembly.

\begin{lstlisting}[
    basicstyle=\scriptsize,
    caption={Assembly code of modified Xalan}, label={lst:xalan_inj_disasm}]
 4ae3f6: e8 e5 1b 00 00          callq  4affe0 
 4ae3fb: e9 00 00 00 00          jmpq   4ae400
 4ae400: bf 30 00 00 00          mov    $0x30,%edi
 4ae405: e8 c6 2c 12 00          callq  5d10d0 
 4ae40a: 48 89 85 28 fd ff ff    mov    %rax,-0x2d8(%rbp)
 4ae411: e9 00 00 00 00          jmpq   4ae416
........ (this code below is not disassembled by radare)
*4ae441: 48 8b 85 20 fd ff ff    mov    -0x2e0(%rbp),%rax
*4ae448: 48 89 45 c8             mov    %rax,-0x38(%rbp)
*4ae44c: c7 05 02 e7 5a 00 43    movl   $0x434343,0x5ae702(%rip)
*4ae453: 43 43 00
*4ae456: e8 15 ed ff ff          callq  xercesc::bad
*4ae45b: e9 00 00 00 00          jmpq   4ae460
......
 4aff9e: 48 8d 7d d8             lea    -0x28(%rbp),%rdi
 4affa2: e8 29 9a 04 00          callq  4f99d0
 4affa7: e9 13 00 00 00          jmpq   4affbf
\end{lstlisting}

\section{Control flow instruction error}
\label{app:controlflow}

We give some examples of control flow errors classified into direct/indirect and conditional/unconditional.

\subsection{Direct}
% \balance
\label{app:direct}
We found that an instruction called by a direct control transfer instruction can be missed. The example here shows that \ida can also miss the instruction that is called through direct call. This example is in {\tt Xalan} 2017 compiled with \verb+gcc -O0+. \ida does not disassemble instruction at address 0xe538d, which is directly called from address 0xe53e7. Listing \ref{lst:ida_direct_err} showing the direct call instruction at address 0xe53e7, which \ida disassembled, and instructions at address 0xe538d which did not disassembled.

\begin{lstlisting}[
    basicstyle=\footnotesize,
    caption={\ida missed }, label={lst:ida_direct_err}]
e538d: ..... (not disassembled) 
e53e7: call   e538d (disassembled)
\end{lstlisting}

\subsection{Conditional Branch}
\label{app:cbr}
We found that the target and fall-through of a conditional branch instruction can be missed by the disassembler. The example below was disassembled by \angr for {\tt omnetp} 2017 compiled with \verb+gcc -O3 LTO+ and stripped. 
Listing~\ref{lst:cbr_example} below shows the ground truth disassembly and Listing~\ref{lst:cbrerr_example} the \angr disassembly.
\angr chooses to skip address 0x183162 and starts disassembling at address 0x183170. 
We remark that due to branch at 0x183160, a recursive descent disassembler may not choose to consider bytes from 0x183162 to be in the code depending on the analysis of the rest of the code.

\begin{lstlisting}[
basicstyle=\footnotesize,
caption={Ground truth for omnetp}, label={lst:cbr_example}]
183160: 7f 26                   jg     183188 
183162: 48 8d 05 3b 60 03 00    lea    0x3603b(%rip),%rax
183169: 85 f6                   test   %esi,%esi
18316b: 74 2e                   je     18319b
18316d: 83 fe 01                cmp    $0x1,%esi
183170: ba 00 00 00 00          mov    $0x0,%edx
\end{lstlisting}

\begin{lstlisting}[
basicstyle=\footnotesize,
caption={Angr result for omnetp}, label={lst:cbrerr_example}]
183160: 7f 26             jg 183188 
183165: 3b 60 03          cmpl $3(%rax), %esp
183170: ba 00 00 00 00    mov    $0x0,%edx
\end{lstlisting}

\subsection{Return}
\label{app:return}
When disassembling {\tt cactusADM} 2006 compiled with {\tt clang -O3 LTO} with \ghidra, it was observed that the target for the return instruction was not disassembled, shown below in Listing~\ref{lst:reterr_example}.

\begin{lstlisting}[
basicstyle=\footnotesize,
caption={cactusADM incorrect disassembly}, label={lst:reterr_example}]
42a14e: 31 c0           xor        eax, eax
42a150: e8 0b 1d 00 00  callq      CCTK_VWarn
42a155: ......          (not disassembled)
\end{lstlisting}

\ghidra assumes after its analysis that function {\small \verb+CCTK_VWarn()+} does not return, i.e., it behaves as if the attribute {\footnotesize \verb+_attribute__ ((noreturn))+} is specified on that function. 
Hence, instructions after the call instruction are not disassembled. However, \tool confirms that the instruction at address {\tt 0x42a155} is executed.

\subsection{Indirect}
\label{app:indirect}
Recursive descent disassemblers have heuristics to guess the possible targets of indirect control flow. As the heuristics are not guaranteed correct, we can expect errors to occur during static analysis.
For instance, \ghidra failed to identify some instances of indirect control flow in the \xz 2017 program compiled with \verb+gcc LTO+. Listing~\ref{lst:indir_example} below gives an example. The instruction at address 0xca9c is an indirect jump. Our control flow explanation showed that there were several target addresses, such as \{0xcaa0, 0xcdc0, 0xce90\} but \ghidra failed to disassemble any of these basic blocks.

\begin{lstlisting}[
basicstyle=\footnotesize,
caption={XZ indirect jump}, label={lst:indir_example}]
ca8e: 48 8d 1d 9f 19 01 00    lea    0x1199f(%rip),%rbx
ca95: 48 63 0c 8b             movslq (%rbx,%rcx,4),%rcx
ca99: 48 01 d9                add    %rbx,%rcx
ca9c: 3e ff e1                jmpq *%rcx
\end{lstlisting}

Apart from that, most of the errors caused by indirect control flow in \angr happen in the PLT section, as shown in Listing~\ref{lst:plt_clang}.

\section{Wrongly Disassembled Instructions} \label{appendix:wrong_disasm}
Missing instructions can cause disassemblers to produce incorrect disassembly. An example of this is observed when \angr disassembles \perlbench 2006 compiled with \verb+gcc -O3+. The disassembly misses the fallthrough of a conditional branch, which subsequently leads to the incorrect disassembly of the following instruction. Listing \ref{lst:wrong_disasm_ex} shows the correct disassembly, and Listing \ref{lst:wrong_disasm} shows the disassembly result yielded by \angr.

\begin{lstlisting}[
basicstyle=\footnotesize,
caption={Ground truth for perlbench}, label={lst:wrong_disasm_ex}]
a20a0: 0f 84 1a 09 00 00       je a29c0 
a20a6: 41 8b 56 68             movl 0x68(%r14), %edx
a20aa: 85 d2                   testl %edx, %edx
a20ac: 0f 84 de 06 00 00       je a2790
a20b2: 41 f7 06 00 40 00 00    testl $0x4000, (%r14)
\end{lstlisting}

\begin{lstlisting}[
basicstyle=\footnotesize,
caption={Angr result for perlbench}, label={lst:wrong_disasm}]
a20a0: 0f 84 1a 09 00 00      je 0x4a29c0
a20a8: 56                     pushq %rsi
a20a9: 68 85 d2 0f 84         pushq $0x840fd285
a20ae: de 06                  fiadds (%rsi)
a20b0: 00 00                  addb %al, (%rax)
\end{lstlisting}

\end{document}